\documentstyle[12pt]{article}

\newcommand{\beq}{\begin{equation}}
\newcommand{\eeq}{\end{equation}}
\newcommand{\beqa}{\begin{eqnarray}}
\newcommand{\eeqa}{\end{eqnarray}}
\newcommand{\nn}{\nonumber}
\newcommand{\eq}[1]{(\ref{#1})}

\newcommand{\ddG}{G^{(1)} {\mskip-26mu\raise10pt
   \vbox{\kern7pt\hbox{.}}\mskip1mu\raise5pt\hbox{$'$}\mskip1mu}}
\newcommand{\ddGcal}{{\cal G}^{(1)} {\mskip-25mu\raise10pt
\vbox{\kern7pt\hbox{.}} \mskip1mu\raise5pt\hbox{$'$}\mskip1mu}}
\newcommand{\ra}{\rightarrow}
\newcommand{\tr}{{\rm Tr}}
\newcommand{\eps}{\epsilon}
\newcommand{\NP}[1]{ {\it Nucl.{}~Phys.} {\bf #1}}
\newcommand{\PL}[1]{ {\it Phys.{}~Lett.} {\bf #1}}

\newcommand{\PR}[1]{ {\it Phys.{}~Rev.} {\bf #1}}
\newcommand{\PRL}[1]{ {\it Phys.{}~Rev.{}~Lett.} {\bf #1}}

\begin{document}
\begin{titlepage}
\setcounter{page}{0}
\begin{flushright}
NBI-HE-96-54\\
hep-th/9610064\\
\end{flushright}

\vspace{15mm}
\begin{center}
{\Large  Pinched Gluon Vertex Operator in Super Worldline Formalism} 
\vspace{20mm}

{\large Haru-Tada Sato   
\footnote{sato@nbi.dk}}\\
{\em Niels Bohr Institute, University of Copenhagen\\
     Blegdamsvej 17, DK-2100 Copenhagen, Denmark}\\
\end{center}
\vspace{7mm}

\begin{abstract}
We reformulate, using super worldline formalism, the pinched gluon 
vertex operator proposed by Strassler. The pinched vertex operator 
turns out to be the product of two gluon vertex operators with the 
insertion of $\delta$-function which makes the super distances 
between them zero. Thus the pinch procedures turn out to be nothing 
but the insertions of $\delta$-function. Applying our formulation 
to two-loop diagrams which are the QED correction to gluon scatterings 
via a single spinor loop, with the QED charge $e$ being replaced by 
the strong coupling $g$, we show various formulae on pinched 
$N$-point functions.                                               
\end{abstract}

\vspace{1cm}
\begin{flushleft}
\end{flushleft}
\end{titlepage}
\newpage
\renewcommand{\thefootnote}{\arabic{footnote}}
\section{Introduction}
\setcounter{equation}{0}
\indent

The worldline formalism of field theory has become an important 
concept in the relevance to the Bern-Kosower rules, \cite{BK} which 
provide a simple reorganization of one-loop Feynman amplitudes in 
Yang-Mills gauge theory (see \cite{BKrev} for a review). Roughly 
speaking, we have two advantages in this formalism. In the first 
place, one is a technical merit to get rid of tedious and 
extravagant Feynman rule calculation. For example, 5-point gluon 
scattering amplitude was calculated through the Bern-Kosower rules 
\cite{BDK}. The simplified calculation method seems to be promising 
to obtain further complicated results such as graviton scatterings 
\cite{BDS} and multi-loop generalization \cite{Kaj}. Although the 
worldline formalism only generates the effective action (the sum of 
1PI graphs) and not the full scattering amplitudes, it certainly 
plays an important role to make a connection between particle and 
string theories. 
Another one is a clarification of underlying physical or mathematical 
structure and concept, for example, worldline supersymmetry, which 
enables us to express gluon amplitudes as correlation functions of 
worldline superfields. These interesting properties are strongly 
related to string and conformal field theories and thus provide a 
phenomenological motivation of studying lower dimensional field 
theories. None of these points can be understood through standard 
Feynman rule calculation.

There are already many one-loop studies of worldline formalism 
\cite{str}-\cite{const}. However the results of this approach 
concerning two-loop theories \cite{phi3}-\cite{PDV} have not been so 
much accumulated as those of one-loop theories. Only simple 
theories, $\phi^3$-theory~\cite{phi3}, spinor QED~\cite{qed}, 
$\phi^4$-theory and scalar QED ~\cite{sqed}, were investigated. 
Ref.\cite{qed} enables us to calculate, using super-worldline 
techniques, photon scattering amplitudes via single spinor loop with 
one (or more) internal photon insertion(s). Various types of 
multi-loop worldline Green function \cite{phi3},\cite{HTS},\cite{RS} 
are obtained from the viewpoints of string and particle theories. 
The analysis of $\phi^4$-type interaction of \cite{sqed} seems to 
suggest us a clue how the four-point vertices of Yang-Mills theory 
should be handled within the realm of super-worldline formulation. 
The scalar $\phi^3$-theory with an internal color symmetry was also 
analyzed \cite{PDV}, using bosonic string theory, as a preliminary 
for Yang-Mills theory. If we successfully combine all these 
significant results, we can expect a new development toward 
Bern-Kosower-like rules for QCD multi-loop scattering theory. 
This is the present scope beyond the one-loop studies. 

In this paper, we consider, combining two works, two-loop gluon 
scatterings via a single spinor/gluon loop with one internal gluon 
insertion. One is Strassler's work which formulated gluon scatterings 
in one-loop Yang-Mills theory~\cite{str}. In particular, he showed 
that the contribution of gluon self-interaction to the one-loop 
amplitudes can be evaluated by means of inserting a pinched gluon 
vertex operator. Another one is ref.\cite{qed}, in which Schmidt and 
Schubert studied the photon scatterings in two-loop QED using 
super-worldline formulation (Thus, we focus only on the same type 
of diagrams studied by Schmidt-Schubert). The problem of 
multi-loop formulation for the gluon scatterings is to find out 
a method how to evaluate pinched contributions. In the one-loop 
case, it is enough to insert the pinched vertex operators into a 
fundamental loop expression which corresponds to the un-pinched 
function. However, in the higher loop cases, this simple situation 
becomes complicated because one of external lines at a pinched 
vertex should participate in the internal gluon propagator part. For 
this reason, we have to reformulate Strassler's pinch prescription 
into more suitable form to be applicable beyond one-loop order. 
To this end, the super-worldline formalism is very useful.

The contents of this paper are as follows.
First in sect.2, we briefly review how the pinched gluon vertex 
operator works in Strassler's one-loop argument, which is not 
organized into the super-worldline formalism. A special care about 
a $\delta$-function must be taken into account there. In sect.3, we 
reorganize the statements of sect.2 introducing a worldline 
superfield. The pinched gluon vertex operator turns out to be exactly 
the product of two gluon vertex operators which are joined by a 
$\delta$-function at the same vertex on a super-worldline. In this 
formulation, there is no need to give the $\delta$-function the 
special treatment remarked in sect.2. Then in sect.4, applying the 
super-worldline pinch prescription of sect.3 to two-loop $N$-point 
diagrams which are the QED correction to gluon scatterings via a 
single spinor loop with the QED charge $e$ being replaced by the 
strong coupling $g$, we derive several vanishing formulae on the 
main parts (vertex position integrals) of pinched $N$-point 
functions. In sects.5 and 6, we calculate pinched two- and 
three-point functions. In sect. 7, we discuss a connection of our 
pinched functions between one-loop and two-loop cases. In sect. 8, 
we comment on the gluon loop case in short. Conclusion is sect.9.

\section{Notes on Strassler's method}
\setcounter{equation}{0}
\indent                                

It is worth reviewing Strassler's pinch method \cite{str} in order 
to clearly understand the difference between his method and ours, 
which will be explained later. We will be mainly concerned with 
spinor (with bare mass $m$) loop case, since gluon loop case is 
essentially parallel to the spinor case.

The proper $N$-point functions of gluon scatterings with single 
spinor loop in the worldline approach to the Bern-Kosower rules are 
written in the following integral \cite{str} 
\beq
\Gamma_N = -{1\over2}\tr\int_0^\infty{dT\over T}
           e^{-m^2T}{\hat\Gamma}_N.
\label{proper}
\eeq
The value of ${\hat\Gamma}_N$ is given by the expectation value of 
$N$ gluon vertex operators
\beq
{\hat\Gamma}_N = \oint[dx][d\psi] e^{-S_0}
           \prod_{n=1}^N V_n, \qquad V_n \equiv V(k_n,\eps_n),
\label{npoint}
\eeq
where $x(\tau)$ and $\psi(\tau)$ are one-dimensional bosonic and 
fermionic fields, and the gluon vertex operator is 
\beq
V(k,\eps) =(-igT^{a_n})\int_0^T d\tau (\eps_{\mu}{\dot x}^{\mu} 
            +2i\psi^\mu \eps_\mu \psi^\nu k_\nu)\exp[ik_\mu x^\mu].
\label{gluonvertex}
\eeq
The worldline action $S_0$, of which form depends on the particle 
moving around the loop, is 
\beq
 S_0 = \int_0^Td\tau ({1\over4}{\dot x}^2 
       +{1\over2}\psi\cdot{\dot\psi}),     \label{action}
\eeq
where ${\dot x}^2={\dot x}^\mu{\dot x}_\mu$, and ${\dot x}=dx/d\tau$ 
($A\cdot B$ means $A^\mu B_\mu$, and $\mu$ runs over $1,2,\dots D$).
Note also that the minus sign and ${1\over2}$ of the pre-factor 
in eq.\eq{proper} count the statistics and degrees of freedom of the 
particle moving around the loop. For example, in the cases of  
complex boson/ghost loop, the pre-factor must be 
$2\times{1\over2}$ and $-2\times{1\over2}$ (Gluon loop case is 
assigned to ${1\over2}$ exceptionally).

In Strassler's method, the closed path integrals on $x$ and $\psi$ in 
${\hat\Gamma}_N$ are performed introducing additional 
Grassmann integration variables:
\beq
{\hat\Gamma}_N=
{\cal N}\prod_{n=1}^N\int_0^T d\tau_n d\theta_n d{\bar\theta}_n
\exp[ -\int d\tau d\tau' \{
 {1\over2}J^\mu(\tau)G_B(\tau,\tau')J_\mu(\tau')
+{1\over4}\eta^\mu(\tau)G_F(\tau,\tau')\eta_\mu(\tau') \} ],  
\label{unpinchgamma}
\eeq
where
\beq
G_B(\tau_j,\tau_i)=|\tau_j-\tau_i|-{(\tau_j-\tau_i)^2\over T}
\equiv G_B^{ji},
\eeq
\beq
G_F(\tau_j,\tau_i)= \mbox{sign}(\tau_j-\tau_i)
\equiv G_F^{ji},
\eeq
\beq
{\cal N}=\oint[dx][d\psi] e^{-S_0} = 
         2^{D\over2}\left({1\over4\pi T}\right)^{D\over2}
\eeq
and
\beqa
&&J^\mu(\tau)=\sum_{n=1}^N\delta(\tau-\tau_n)({\bar\theta}_n\theta_n
\eps^\mu_n \partial_\tau + ik^\mu_n),\\
&&\eta^\mu(\tau)=\sum_{n=1}^N\delta(\tau-\tau_n)\sqrt{2}
(\theta_n \eps^\mu_n + i{\bar\theta}_n k^\mu_n).
\eeqa

On the other hand, a pinched function is obtained through the 
replacement of a pair of two vertex operators $V_jV_i$ by 
${\cal O}_{ji}$:
\beq
{\hat\Gamma}_N(j,i) \equiv \oint[dx][d\psi] 
      e^{-S_0}\prod_{n\not=i,j}^N V_n{\cal O}_{ji},  \label{gammaji}
\eeq
\beq
{\cal O}_{ji}=(-igT^{a_j})(-igT^{a_i})\int_0^T d\tau_i d\tau_j
\delta(\tau_i-\tau_j) 2 \eps_j\cdot\psi \eps_i\cdot\psi
e^{i(k_i+k_j)\cdot x}.  \label{pinchvertex}
\eeq
Note that the pinched vertex operator ${\cal O}_{ji}$ does never 
look like $V_jV_i$ at this stage. In addition, equality in 
\eq{gammaji} should be understood under an appropriately ordered 
color factor (The positions of $T^{a_j}$ and $T^{a_i}$ are not clear 
in the above formal expression). For simplicity, let us ignore it for 
the moment. A clear explanation will be given in the next section 
along the super-worldline context. Now, similarly as done in 
\eq{unpinchgamma}, we have the path-integrated form of \eq{gammaji}
\beqa
{\hat\Gamma}_N(j,i)
&=&{\cal N}\prod_{n=1}^N\int_0^T d\tau_n d\theta_n d{\bar\theta}_n
           \delta(\tau_i-\tau_j)(k_i\cdot k_j G_F^{ji})^{-1} \nn\\
&&\times\exp[-{\bar\theta}_i{\bar\theta}_j k_i\cdot k_j G_F^{ji}
      -\int d\tau d\tau'
 {1\over2}{\tilde J}^\mu(\tau)G_B(\tau,\tau'){\tilde J}_\mu(\tau')
 \label{pinchgamma} \\
&&\hskip 40pt +{1\over4}{\tilde\eta}^\mu(\tau)G_F(\tau,\tau')
               {\tilde\eta}_\mu(\tau')], \nn     
\eeqa
where
\beqa
&&{\tilde J}^\mu(\tau)=\sum_{n\not=i,j}^N\delta(\tau-\tau_n)
({\bar\theta}_n\theta_n\eps^\mu_n \partial_\tau + ik^\mu_n)
+\sum_{n=i,j}\delta(\tau-\tau_n)ik^\mu_n,\\
&&{\tilde\eta}^\mu(\tau)=\sum_{n\not=i,j}^N\delta(\tau-\tau_n)
\sqrt{2}(\theta_n\eps^\mu_n + i{\bar\theta}_n k^\mu_n)
+\sum_{n=i,j}\delta(\tau-\tau_n)\sqrt{2}\theta_n\eps^\mu_n.
\eeqa
Note that the ${\tilde J}G_B{\tilde J}$ term includes none of 
${\bar\theta}_i$, ${\bar\theta}_j$, $\theta_i$ and $\theta_j$. 
${\tilde\eta}G_F{\tilde\eta}$ does not include ${\bar\theta}_i$ 
and ${\bar\theta}_j$. Hence, as seen from \eq{pinchgamma}, the 
result of Grassmann integrals of the exponential part is exactly 
proportional to $k_i\cdot k_jG_F^{ji}$. Obviously, such terms can be 
alternatively extracted from the un-pinched exponent 
$\exp[-{1\over2}JG_BJ-{1\over4}\eta G_F\eta]$ in \eq{unpinchgamma}, 
in terms of just picking up the terms proportional to 
$k_i\cdot k_jG_F^{ji}$. This means that we have to throw irrelevant 
terms away {\it by hand} from \eq{unpinchgamma}. Otherwise, we have 
to perform $2N$ Grassmann integrals of \eq{pinchgamma}. 
Also note that the $\delta$-function originally included in the pinch 
vertex operator ${\cal O}_{ji}$ should be performed {\it after} 
$(G_F^{ji})^{-1}$ cancels $G_F^{ji}$s coming up from the exponent in 
\eq{pinchgamma}. After all, we must not integrate 
$\delta(\tau_j-\tau_i)$ until we finish two procedures (i) $2N$ 
Grassmann integrals, (ii) cancellation of $(G_F^{ji})^{-1}$. 

As mentioned in \cite{str}, the pinch contribution of two-point 
function ${\hat\Gamma}_2(2,1)$ is zero. The simplest example of 
non-vanishing pinch is the three point function ${\hat\Gamma}_3(2,1)$ 
\beq
{\hat\Gamma}_3(2,1)=
-i{\cal N}\int d\tau_3 d\tau_2 (k_3\cdot\eps_1 \eps_3\cdot\eps_2 
              - k_3\cdot\eps_2 \eps_3\cdot\eps_1)(G_F^{32})^2
              e^{-k_3^2G_B^{23}}.             \label{3pointpinch}
\eeq
One can directly verify, through checking \eq{3pointpinch}, how the 
above attention should be paid in  calculation.
\section{Super worldline formulation}
\setcounter{equation}{0}
\indent 

Let us reformulate the above pinch formulation using super-worldline 
technique \cite{qed}. We will see in this section that the 
super-worldline formulation does not need the careful treatment about 
the $\delta$-function remarked in previous section. This time, we do 
not have to introduce the artificial $2N$ Grassmann integrals, and 
the $\delta$-function can be integrated away immediately. These 
improvements simplify the calculation. Furthermore, differently 
from \eq{pinchvertex}, ${\cal O}_{ji}$ reasonably looks like 
$V_jV_i$ in our formulation.

With the following super worldline notation
\beq
X(\tau,\theta)=x(\tau)+\sqrt{2}\theta\psi(\tau),\qquad
{\cal D}={\partial\over\partial\theta}
                  -\theta{\partial\over\partial\tau},
\eeq
the gluon vertex operator \eq{gluonvertex} and the spinor loop 
action \eq{action} can be re-expressed \cite{qed}
\beqa
&&  S_0=-\int_0^Td\tau d\theta {1\over4} X{\cal D}^3X, \\
&&  V_n=-(-igT^{a_n})\int_0^T d\tau d\theta \eps_n\cdot {\cal D}X 
         \exp[ik_n\cdot X],
\eeqa
and the un-pinched $N$-point function defined in \eq{npoint} reads
\beq
 {\hat\Gamma}_N=\oint[dX]e^{-S_0}\prod_{n=1}^NV_n.  
\label{oneloopgamman}
\eeq
Differently from the argument of sect.2, we apply the Wick 
contraction to the evaluation of this $N$-point function using 
the super-worldline Green function \cite{qed}
\beq
<X^\mu_1X^\nu_2>={\cal N}^{-1}\oint[dX]e^{-S_0}X_1^\mu X_2^\nu 
=-g^{\mu\nu}G(1,2),
\eeq
where
\beq
G(1,2)=G^{12}_B + \theta_1\theta_2 G^{12}_F.
\eeq

Now, the pinched gluon vertex operator \eq{pinchvertex} 
turns out to be 
\beq
{\cal O}_{ji}=(-igT^{a_j})(-igT^{a_i})\int_0^T d\tau_i d\tau_j
d\theta_j d\theta_i \theta_i \theta_j \delta(\tau_i-\tau_j)
\eps_i\cdot{\cal D}X_i \eps_j\cdot{\cal D}X_j
e^{ik_i\cdot X_i +ik_j\cdot X_j },         \label{supervertex}
\eeq
where $X_i$ means $X(\tau_i,\theta_i)$. It is now clear that 
${\cal O}_{ji}$ is created by the product of two vertex operators 
$V_jV_i$. There exists the following equality (denoted by $\sim$) 
between $V_jV_i$ and ${\cal O}_{ji}$ at the level of integrand, 
\beq
{\cal O}_{ji} \sim V_j\theta_j \theta_i \delta(\tau_j-\tau_i)V_i.
\eeq
This means that the insertion of ${\cal O}_{ji}$ is equivalent to 
the insertion of $\theta_j\theta_i\delta(\tau_j-\tau_i)$ between 
$V_j$ and $V_i$ (or ${\cal D}X_j$ and ${\cal D}X_i$). Note that 
$\theta_j\theta_i\delta(\tau_j-\tau_i)$ is the $\delta$-function 
which makes the super-distances $\tau_j-\tau_i+\theta_j\theta_i$ 
zero, and this $\delta$-function coincides with a part of a 
supersymmetric step function \cite{AT}. In the following, we 
consider a fixed color ordering, and the ordering of $V_n$ 
($n=1,2,\cdots N$) should be fixed before inserting the pinch 
$\delta$-function $\theta_j\theta_i\delta(\tau_j-\tau_i)$.
For example, we choose $V_N, V_{N-1}, \cdots, V_1$. Eq.\eq{gammaji} 
then becomes
\beq
{\hat\Gamma}_N(j,i) = \oint [dX] e^{-S_0}  V_N  \cdots V_j \cdots 
          V_i \cdots V_1  \theta_j\theta_i\delta(\tau_j-\tau_i).
\eeq
If one wants to obtain also ${\hat\Gamma}_N(i,j)$, it is enough 
to insert $\theta_i\theta_j\delta(\tau_i-\tau_j)$ and to reverse 
the ordering between $T^{a_j}$ and $T^{a_i}$. The sign of inserted 
$\delta$-function is then reversed, and the sum of these pinched 
functions is naturally accompanied by the structure constant 
produced by the commutator between $T^{a_j}$ and $T^{a_i}$.

Hereafter, we will ignore this kind of color trace structure, since 
this is just a matter of counting factors, and it can be easily 
recovered after getting main expression (integral parts) of 
${\hat\Gamma}_N(j,i)$. By reason of this, keep in mind that the 
vertex operators $V_n$ will be dealt as if commuting quantities.

Let us look how the $\delta$-function insertion, which is equivalent 
to the insertion of super pinch vertex operator \eq{supervertex}, 
works in the two-point function case. The result must be zero as seen 
in sect.2. First, write down ${\hat\Gamma}_2$ using Wick 
contractions
\beqa
{\hat\Gamma}_2
  &=&-{\cal N}\int d\tau_1 d\tau_2 d\theta_1 d\theta_2 \eps^\mu_1 
          \eps^\nu_2<{\cal D}X_\mu^1 {\cal D}X_\nu^2
                     e^{ik_1\cdot X_1} e^{ik_2\cdot X_2}> \nn\\
  &=&-{\cal N}\int d\tau_1 d\tau_2 d\theta_1 d\theta_2 \eps^\mu_1 
         \eps^\nu_2 <e^{ik_1\cdot X_1} e^{ik_2\cdot X_2}> \nn \\
  & &\times [{\cal D}_1{\cal D}_2<X^\mu_1X^\nu_2>-
  {\cal D}_1<X^\mu_1k_2\cdot X_2>{\cal D}_2<X^\nu_2k_1\cdot X_1>],
\label{twopoint}
\eeqa
and substitute
\beqa
&&{\cal D}_1{\cal D}_2<X^\mu_1X^\nu_2>= g^{\mu\nu}
              (G^{12}_F +\theta_1 \theta_2 {\ddot G}^{12}_B), 
\label{formula1} \\
&&{\cal D}_1<X^\mu_1 k_2\cdot X_2>=k^\mu_2(\theta_1{\dot G}^{12}_B
                 -\theta_2 G^{12}_F), 
\label{formula2} \\ 
&&<e^{ik_1\cdot X_1}e^{ik_2\cdot X_2}>=e^{k_1\cdot k_2 G(1,2)},
\eeqa
where ${\dot G}_B$ and ${\ddot G}_B$ mean the first and second 
derivatives w.r.t. the first argument of $G_B$.
Following the method explained above, the pinched two-point function 
${\hat\Gamma}_2(2,1)$ can be obtained by inserting 
$\theta_2\theta_1\delta(\tau_2-\tau_1)$ into eq.\eq{twopoint}.
We then immediately see
\beq
{\hat\Gamma}_2(2,1)=\oint[dX]e^{-S_0}V_2V_1\theta_2\theta_1
\delta(\tau_2-\tau_1)=0,
\eeq
while getting the un-pinched function as well 
\beqa
{\hat\Gamma}_2
&=&{\cal N}\int d\tau_1 d\tau_2 e^{k_1\cdot k_2G^{12}_B}[
\eps_1\cdot\eps_2 {\ddot G}^{12}_B + \eps_1\cdot k_2 \eps_2\cdot k_1
({\dot G}^{12}_B)^2 \nn \\
& &+ (\eps_1\cdot\eps_2 k_1\cdot k_2-\eps_1\cdot k_2\eps_2\cdot k_1)
(G^{12}_F)^2],
\eeqa
which coincides with equation (3.28) of ref.\cite{str}. It is 
clear that the insertion of $\theta_2\theta_1$ removes irrelevant 
terms from \eq{twopoint} and that $\delta(\tau_2-\tau_1)$ can be 
integrated right at the moment when it is inserted. 
(A $\delta$-function multiplied by a sign function, i.e. 
$G_F^{12}\delta(\tau_2-\tau_1)$, does not have a well-defined value 
when $\tau_1=\tau_2$. However, if one considers the $\delta$-function 
to be even, then the result follows.) 
These are the different points 
from the method of sect.2 and make calculation simple.

Next, let us check whether our method works in a non-trivial 
example, say ${\hat\Gamma}_3(2,1)$. Writing down the un-pinched 
three-point function
\beqa
{\hat\Gamma}_3
&=&{\cal N}\int d\tau_3 d\tau_2 d\tau_1 d\theta_3 d\theta_2 d\theta_1
\eps^\mu_3\eps^\nu_2\eps^\rho_1<{\cal D}X_\mu^3{\cal D}X_\nu^2
{\cal D}X_\rho^1 \prod_{j=1}^3 e^{ik_j\cdot X_j}> \nn\\
&=&
{\cal N}\int d\tau_3 d\tau_2 d\tau_1 d\theta_3 d\theta_2 d\theta_1
\eps^\mu_3\eps^\nu_2\eps^\rho_1\exp[\sum_{i<j}k_i\cdot k_jG(i,j)]
K_{\mu\nu\rho},   
\eeqa
where
\beqa
K^{\mu\nu\rho}&=& {\cal D}_3{\cal D}_2<X^\mu_3X^\nu_2>
             {\cal D}_1<X^\rho_1\sum_{j=1}^3ik_j\cdot X_j> \nn \\
&-&{\cal D}_3{\cal D}_1<X^\mu_3X^\rho_1>
   {\cal D}_2<X^\nu_2\sum_{j=1}^3ik_j\cdot X_j>\label{Kmunurho}\\
&+&{\cal D}_2{\cal D}_1<X^\nu_2X^\rho_1>
               {\cal D}_3<X^\mu_3\sum_{j=1}^3ik_j\cdot X_j>, \nn
\eeqa                                                
we insert $\theta_2\theta_1\delta(\tau_2-\tau_1)$ according to 
the integrand equality ${\hat\Gamma}_3(2,1)\sim{\hat\Gamma}_3
\theta_2\theta_1\delta(\tau_2-\tau_1)$. Obviously, the third term of 
$K_{\mu\nu\rho}$ vanishes after this pinch procedure 
(see eq.\eq{formula1}). Applying the formulae 
\eq{formula1} and \eq{formula2} to $K_{\mu\nu\rho}$ and integrating 
w.r.t. $\tau_1$ and all $\theta_i$, we arrive at 
\beqa
{\hat\Gamma}_3(2,1) &=& \oint[dX]e^{-S_0}V_3V_2V_1\theta_2\theta_1
\delta(\tau_2-\tau_1) \nn \\
&=& {\cal N}\int d\tau_3 d\tau_2 \eps^\mu_3\eps^\nu_2\eps^\rho_1
i(-g_{\mu\nu}k^\rho_3  + g_{\mu\rho}k^\nu_3)(G^{32}_F)^2
e^{-k_3^2 G_B^{23}}.     \label{3pointpinch2}
\eeqa
This coincides with \eq{3pointpinch}.

We can calculate pinched 4-point functions in the same way. There 
are two types: the single-pinch type 
\beqa
{\hat\Gamma}_4(2,1)&=& {\cal N}\int d\tau_4 d\tau_3 d\tau_1
\exp[k_4\cdot k_3 G_B^{43} + k_4\cdot(k_1+k_2) G_B^{41} 
              + k_3\cdot(k_1+k_2) G_B^{31} ]  \nn \\
&\times& [\, G_F^{43}G_F^{31}G_F^{14}(
\eps_4\cdot\eps_3 \eps_2\cdot k_3 \eps_1\cdot k_4
-\eps_4\cdot\eps_3 \eps_2\cdot k_4 \eps_1\cdot k_3
+\eps_4\cdot\eps_2 \eps_3\cdot k_4 \eps_1\cdot k_3 \nn \\
& &\hskip 50pt -\eps_4\cdot\eps_1 \eps_3\cdot k_4 \eps_2\cdot k_3 
-\eps_3\cdot\eps_2 \eps_4\cdot k_3 \eps_1\cdot k_4
+\eps_3\cdot\eps_1 \eps_4\cdot k_3 \eps_2\cdot k_4 ) \nn\\
& &\hskip5pt+ {\dot G}_B^{31}(G_F^{41})^2\{
\eps_4\cdot\eps_2 \eps_3\cdot (k_1+k_2) \eps_1\cdot k_4
-\eps_4\cdot\eps_1 \eps_3\cdot (k_1+k_2) \eps_2\cdot k_4\} 
\label{4pointpinch21} \\
& &\hskip5pt+ {\dot G}_B^{41}(G_F^{31})^2\{
\eps_3\cdot\eps_2 \eps_4\cdot (k_1+k_2) \eps_1\cdot k_3
-\eps_3\cdot\eps_1 \eps_4\cdot (k_1+k_2) \eps_2\cdot k_3\}\,],  \nn
\eeqa
and the double-pinch type
\beqa
{\hat\Gamma}_4(4,3|2,1) &\equiv& \oint[dX]e^{-S_0}V_4V_3V_2V_1
                      \theta_4\theta_3\theta_2\theta_1
               \delta(\tau_4-\tau_3)\delta(\tau_2-\tau_1) \nn\\
&=& {\cal N}\int d\tau_3 d\tau_1
\exp[(k_3+k_4)\cdot(k_1+k_2)G_B^{31}] \nn \\
& &\hskip20pt\times(\eps_4\cdot\eps_1 \eps_3\cdot\eps_2 -
\eps_4\cdot\eps_2 \eps_3\cdot\eps_1) (G_F^{31})^2.
\label{4pointpinch4321}
\eeqa

Here is one remark. If we consider scalar vertex operators 
discarding polarization vectors $V=\int d\tau d\theta e^{ikX}$, 
where super field expression may be kept because fermionic integral 
becomes just a normalization, we can recover a four-point function 
of $\phi^4$-theory from \eq{4pointpinch4321}, 
\beq
{\hat\Gamma}_4 = {\cal N}\int d\tau_3 d\tau_1
                \exp[(k_3+k_4)\cdot(k_1+k_2)G_B^{31}].
\eeq

\section{Two-loop pinch formulas}
\setcounter{equation}{0}
\indent 

The super worldline formulation of multi-loop $N$-photon amplitudes 
are discussed in \cite{qed} for the set of diagrams that $p$ photon 
propagators are inserted into the spinor loop. For this type of 
$(p+1)$-loop $N$-point function, we have only to evaluate 
$(N+2p)$-point correlation function. Here, we confine ourselves to 
the two-loop case $p=1$ because we do not have to make our situation 
complex. 

First, let us recall that 
those two-loop $N$-point functions in QED are proportional to the 
following integral \cite{qed}
\beq
\Gamma_N^{(1)}\equiv\int{dT\over T}e^{-m^2T}\int d{\bar T}
(4\pi{\bar T})^{-D\over2}{\hat\Gamma}_N^{(1)},
\eeq 
where an appropriate pre-factor associated to the theory is again 
dropped. ${\hat\Gamma}_N^{(1)}$ is given by the integrals of 
$(N+2)$-point correlator
\beq
{\hat\Gamma}^{(1)}_N = 
\int d\tau_a d\tau_b d\theta_a d\theta_b {\cal N}_1
<{\cal D}X_b\cdot {\cal D}X_a\prod_{n=1}^NV_n>_{(1)},
\label{twoloop}
\eeq
where the path integral normalization ${\cal N}_1$ is 
\beqa
{\cal N}_1 &=& \oint [dX] e^{-S_0} \exp 
               \left[-{(X_a-X_b)^2\over4{\bar T}}\right] \nn\\
&=& 2^{D\over2}(4\pi T)^{-D\over2}(1+{1\over{\bar T}}G(a,b))^{-D/2}.
\eeqa
With the use of Wick contraction, the correlator $<\cdots>_{(1)}$ 
can be decomposed into two-point correlators, i.e., two-loop 
worldline Green functions, 
\beqa
<X^\mu_1X^\nu_2>_{(1)}&=&{\cal N}^{-1}_1\oint[dX] e^{-S_0}\exp
\left[-{(X_a-X_b)^2\over4{\bar T}}\right]X^\mu_1X^\nu_2 \nn \\
&=& -g^{\mu\nu} {\cal G}^{(1)}(1,2).    \label{propX}
\eeqa
These equations are derived by connecting a pair of external photon 
lines through one internal photon propagator. After Wick contracting, 
the $N$-point function ${\hat\Gamma}_N^{(1)}$ generally takes the 
following form
\beq
\eps_1^\mu\dots\eps_N^\nu K_{\mu\dots\nu}^{(1)}
\exp[{1\over2}\sum_{i,j=1}^N k_i\cdot k_j 
                        {\cal G}^{(1)}(i,j)], \label{gammaform}
\eeq
where $K^{(1)}_{\mu\dots\nu}$ consists of possible contraction terms 
among $X$ fields, such as \eq{Kmunurho}. 
A pinched $N$-point function can be obtained through inserting a 
pinch $\delta$-function into ${\hat\Gamma}_N^{(1)}$. 

Before considering pinch situation, we have to remark on the 
ambiguity problem of multi-loop worldline Green functions. 
For example, two-loop worldline Green functions are known as the 
following two forms \cite{phi3},\cite{qed} 
\beq
{\tilde G}^{(1)}(x,y)=G(x,y)+
   {1\over2}{(G(x,a)-G(x,b))(G(y,a)-G(y,b))\over{\bar T}+G(a,b)},
\label{Gofss}
\eeq
or
\beq
 G^{(1)}(x,y)=G(x,y)-{1\over4}{(G(x,a)-G(x,b)-G(a,y)+G(b,y))^2
               \over {\bar T}+G(a,b)}. \label{Gfromstring}
\eeq
The first form was used in deriving the two-loop QED 
$\beta$-function \cite{qed}. The latter form can be derived from a  
pinching ($\alpha'\ra0$) limit of closed string theory \cite{RS}. 
Both are related by the relation
\beq
G^{(1)}(x,y)={\tilde G}^{(1)}(x,y)-{1\over2}{\tilde G}^{(1)}(x,x)
             -{1\over2}{\tilde G}^{(1)}(y,y).
\eeq
This modification is harmless to the exponential part of 
\eq{gammaform}, i.e.
\beq
\exp[{1\over2}\sum_{i,j}k_i\cdot k_jG^{(1)}(i,j)]
=\exp[{1\over2}\sum_{i,j}k_i\cdot k_j{\tilde G}^{(1)}(i,j)],
\eeq
because of momentum conservation concerning external legs. However, 
it is not clear to the Wick contraction parts 
$K^{(1)}_{\mu\dots\nu}$, whether or not both \eq{Gofss} and 
\eq{Gfromstring} give the same results mutually.

Similarly to the one-loop case, pinched contributions can be 
evaluated by means of replacing a pair of $V_jV_i$ with the pinched 
vertex operator ${\cal O}_{ji}$, namely, by insertion of 
$\theta_j\theta_i\delta(\tau_j-\tau_i)$ between ${\cal D}X_j$ and 
${\cal D}X_i$. In accordance with a color ordering, we again fix the 
ordering of all ${\cal D}X_n$ ($n=a,b,1,2,\dots N$) before we insert 
the pinched vertex operator i.e. the $\delta$-function 
$\theta_j\theta_i\delta(\tau_j-\tau_i)$. The ordering we choose 
here is $a,1,2,\dots N,b$ from right to left. Thus (for $j>i$),
\beqa
{\hat\Gamma}^{(1)}_N(j,i)
&=&\int d\tau_a d\tau_b d\theta_a d\theta_b{\cal N}_1 g_{\mu\nu}
             <{\cal D}X^\mu_b\prod_{n=1}^NV_n {\cal D}X^\nu_a>_{(1)} 
                      \theta_j\theta_i\delta(\tau_j-\tau_i)\nn\\
  &\sim& {\hat\Gamma}^{(1)}_N\theta_j\theta_i\delta(\tau_j-\tau_i).
\label{pinchdef}
\eeqa

Note that the pinch of both edges of internal gluon line becomes zero 
irrespectively of the number of external gluon legs
\beq
              {\hat \Gamma}^{(1)}_N(b,a)=0,       \label{pinchab}
\eeq
in which $\theta_b\theta_a$ should be inserted between ${\cal D}X_b$ 
and ${\cal D}X_a$. This can be proved by direct calculation for each 
choice of $G^{(1)}$ or ${\tilde G}^{(1)}$ ($N=0,1$ cases are 
checked), however the following proof is simple for generic $N$. 
Interchanging the integration variables carrying $a$ and $b$ in
\beq
{\hat\Gamma}^{(1)}_N(b,a)=\int d\tau_a d\tau_b d\theta_a d\theta_b
   {\cal N}_1<{\cal D}X_b\cdot{\cal D}X_a\prod_{n=1}^N V_n>_{(1)}
        \theta_b\theta_a\delta(\tau_a-\tau_b),
\eeq
and anti-commuting ${\cal D}X_a$ with ${\cal D}X_b$, we see RHS of 
the above equation becomes $-{\hat\Gamma}^{(1)}_N(b,a)$. Therefore 
${\hat\Gamma}^{(1)}_N(b,a)=0$. In particular, 
${\hat\Gamma}^{(1)}_0(b,a)=0$ is naturally understood if we notice 
two facts that the zero-point function ${\hat\Gamma}^{(1)}_0$ 
originally corresponds to the one-loop two-point function 
${\hat\Gamma}_2$ and that the one-loop pinched two-point function 
${\hat\Gamma}_2(2,1)$ is zero. More general correspondence between 
two-loop $N$-point functions and one-loop $(N+2)$-point functions 
will be discussed in later section. 

In the same way as the proof of \eq{pinchab}, we obtain another 
pinch formula between internal and external gluon lines
\beq
{\hat\Gamma}^{(1)}_N(b,n) + {\hat\Gamma}^{(1)}_N(n,a)=0. 
\label{pinchsum}
\eeq
This formula reduces to \eq{pinchab} if we put $n=a$ or $b$. 

We put one remark. When directly checking \eq{pinchsum}, there is a 
subtle difference in calculation between $G^{(1)}$ and 
${\tilde G}^{(1)}$ (although the final result is independent of this 
choice). For example, consider $\Gamma^{(1)}_1$. On the one hand for 
${\tilde G}^{(1)}$, ${\hat\Gamma}^{(1)}_1(b,1)\not=0$ by itself. 
It is cancelled by 
\beq
{\hat \Gamma}^{(1)}(1,a)=
{1\over2}i\eps\cdot k(D-1)\int d\tau_a d\tau_b {\cal N}_1^B
(G_F^{ab})^2{G_B^{ab}\over {\bar T}+G_B^{ab} },
\eeq
where ${\cal N}_1^B$ consists of only bosonic Green function
\beq
{\cal N}_1^B=2^{D\over2}(4\pi T)^{-D/2}
             (1+{1\over{\bar T}}G_B^{ab})^{-D/2}.
\eeq
On the other hand for $G^{(1)}$,
\beq 
{\hat\Gamma}^{(1)}_1(b,1)={\hat\Gamma}^{(1)}_1(1,a)=0. \label{star}
\eeq

Note that the interchange symmetry of integration variables 
$a\leftrightarrow b$ makes the proofs of \eq{pinchab} and 
\eq{pinchsum} simple. For this reason, their explicit check of  
choice-independence of ${\cal G}^{(1)}$ has not been needed. In the 
same way, we can show the similar formulae for arbitrary number of 
pinch pairs ($n_i,m_i\not=a,b$)
\beqa
&&{\hat\Gamma}_N^{(1)}(b,a|n_1,m_1|\cdots|n_k,m_k)=0,\label{gpab}\\
&&{\hat\Gamma}_N^{(1)}(b,n|n_1,m_1|\cdots|n_k,m_k)
  +{\hat\Gamma}_N^{(1)}(n,a|n_1,m_1|\cdots|n_k,m_k)=0,\label{gpsum}
\eeqa
where the multi-pinch functions are defined by 
\beq
{\hat\Gamma}_N^{(1)}(n_1,m_1|\cdots|n_k,m_k)
\sim{\hat\Gamma}_N^{(1)}\theta_{n_1}\theta_{m_1}
\cdots\theta_{n_k}\theta_{m_k}
\prod_{i=1}^k \delta(\tau_{n_i}-\tau_{m_i}).
\eeq
In general cases, there is no such useful interchange symmetry. 
We must perform and compare direct calculation in each choice of 
${\cal G}^{(1)}$. Eq.\eq{gpsum} is due to the abelian approximation 
(the color structure at the $a$- and $b$- vertices will destroy 
the relation), and it is still unclear whether or not the equation 
comes from gauge invariance. 

These pinch formulae, \eq{pinchab}, \eq{pinchsum}, 
\eq{gpab} and \eq{gpsum}, tell us that (sum of) all the internal 
`gluon' pinch contributions may be ignored, with the exception of the 
diagrams where two edges of an internal `gluon' are pinched to 
distinct external gluon lines like
\beq
{\hat\Gamma}^{(1)}_N(b,n|m,a|n_1,m_1|\cdots|n_k,m_k). \label{bnma}
\eeq
Hence in our particular situation of QED correction, we have only to 
gather external gluon pinch functions
${\hat\Gamma}^{(1)}_N(n,m|\cdots)$ as well as the types of \eq{bnma}. 
In the following sections, we calculate 2- and 3-point functions 
where double-pinch functions of the type \eq{bnma} appear.

\section{Pinched two-point functions}
\setcounter{equation}{0}
\indent 

In this section, we verify, calculating pinched two-point functions, 
the consistency between ${\tilde G}^{(1)}$ and $G^{(1)}$.  
As mentioned at the end of sect.4, the candidates of non-vanishing 
pinched contributions are the following single-pinch function
\beq
{\hat\Gamma}^{(1)}_2(2,1)=\int d\tau_a d\tau_b d\theta_a d\theta_b 
{\cal N}_1g_{\mu\nu}<{\cal D}X^\mu_b V_2 V_1 {\cal D}X^\nu_a>_{(1)}
\theta_2\theta_1 \delta(\tau_2-\tau_1),  \label{single}
\eeq
and the double-pinch function of the type \eq{bnma} 
\beq
{\hat\Gamma}^{(1)}_2(b,2|1,a)
=\int d\tau_a d\tau_b d\theta_a d\theta_b 
g_{\mu\nu}{\cal N}_1<{\cal D}X^\mu_b V_2 V_1 {\cal D}X^\nu_a>_{(1)}
\theta_b \theta_2 \theta_1 \theta_a
\delta(\tau_a-\tau_1) \delta(\tau_b-\tau_2).
\label{double}
\eeq
Although $V_1$ and $V_2$ commute with each other (up to color 
factors), this does not mean the interchange symmetry of integration 
variables $1\leftrightarrow2$. Hence there is no simple analysis 
compared to \eq{pinchab} and \eq{pinchsum} where the interchange 
symmetry $a\leftrightarrow b$ was useful for their proofs.

We perform direct calculations. First, we write down the un-pinched 
function
\beq
{\hat\Gamma}_2^{(1)} = \int d\tau_a d\tau_b d\tau_1 d\tau_2 
d\theta_a d\theta_b d\theta_1 d\theta_2\eps_1^\mu \eps_2^\nu 
{\cal N}_1<{\cal D}X_b\cdot{\cal D}X_a{\cal D}X_2^\nu{\cal D}X_1^\mu
e^{ik_1\cdot X_1}e^{ik_2\cdot X_2}>_{(1)}.                 
\eeq
Using the Wick contraction, the sixfold correlator takes the 
following form 
\beq
<{\cal D}X_b\cdot{\cal D}X_a{\cal D}X_2^\nu{\cal D}X_1^\mu
e^{ik_1\cdot X_1}e^{ik_2\cdot X_2}>_{(1)}
=K^{(1)}_{\mu\nu}\exp[k_1\cdot k_2 {\cal G}^{(1)}(1,2)],
\eeq
and $\eps_1^\mu\eps_2^\nu K^{(1)}_{\mu\nu}$ is given by the following 
15 terms
\beqa
\eps_1^\mu \eps_2^\nu K^{(1)}_{\mu\nu}  
&=&\eps_1\cdot\eps_2 [ 
 D {\ddGcal}_{ab}{\ddGcal}_{12} 
 - {\ddGcal}_{a1}{\ddGcal}_{b2} 
 + {\ddGcal}_{a2}{\ddGcal}_{b1} ] \nn \\
& &+\eps_1\cdot\eps_2 k_1\cdot k_2 [
-  {\dot{\cal G}}^{(1)}_{a1}{\ddGcal}_{12}{\dot{\cal G}}^{(1)}_{b2}
+  {\dot{\cal G}}^{(1)}_{b1}{\ddGcal}_{12}{\dot{\cal G}}^{(1)}_{a2}]
\nn\\
& &+\eps_1\cdot k_2 \eps_2\cdot k_1 [ 
   {\dot{\cal G}}^{(1)}_{a1}{\ddGcal}_{b2}{\dot{\cal G}}^{(1)}_{12}
-  {\dot{\cal G}}^{(1)}_{b1}{\ddGcal}_{a2}{\dot{\cal G}}^{(1)}_{12}
+D {\dot{\cal G}}^{(1)}_{21}{\ddGcal}_{ab}{\dot{\cal G}}^{(1)}_{12}
\nn\\
& &\hskip 65pt
-  {\dot{\cal G}}^{(1)}_{21}{\ddGcal}_{a1}{\dot{\cal G}}^{(1)}_{b2}
+  {\dot{\cal G}}^{(1)}_{21}{\ddGcal}_{b1}{\dot{\cal G}}^{(1)}_{a2} ] 
\label{kernel}\\
& &+ \eps_1\cdot k_1 \eps_2\cdot k_2 [ 
-  {\dot{\cal G}}^{(1)}_{a1}{\ddGcal}_{b1}{\dot{\cal G}}^{(1)}_{22}
+  {\dot{\cal G}}^{(1)}_{b1}{\ddGcal}_{a1}{\dot{\cal G}}^{(1)}_{22}
-D {\dot{\cal G}}^{(1)}_{11}{\ddGcal}_{ab}{\dot{\cal G}}^{(1)}_{22}
\nn\\
& &\hskip 65pt
+ {\dot{\cal G}}^{(1)}_{11}{\ddGcal}_{a2}{\dot{\cal G}}^{(1)}_{b2}
- {\dot{\cal G}}^{(1)}_{11}{\ddGcal}_{b2}{\dot{\cal G}}^{(1)}_{a2}], 
\nn
\eeqa
where
\beq
{\dot{\cal G}}^{(1)}_{ij} = {\cal D}_i {\cal G}^{(1)}(i,j),
\qquad
{\ddGcal}_{ij} = {\cal D}_i{\cal D}_j{\cal G}^{(1)}(i,j).
\eeq
To estimate \eq{single} and \eq{double}, we insert $\theta_2\theta_1$ 
at least, and we thereby discard the terms proportional to $\theta_1$ 
or $\theta_2$ contained in the above 15 terms. Then pick up the terms 
proportional to 1 for ${\hat\Gamma}_2^{(1)}(b,2|1,a)$, and similarly 
those proportional to $\theta_a\theta_b$ for 
${\hat\Gamma}_2^{(1)}(2,1)$. Note that the exponent 
$k_1\cdot k_2 {\cal G}^{(1)}_{12}$ does not contribute to these 
pinched functions, but it does to ${\hat\Gamma}_2^{(1)}(1,a)$.
Note also that ${\ddGcal}_{xy}$ does not depend on the choice whether 
$G^{(1)}$ or ${\tilde G}^{(1)}$.

Now, let us consider ${\hat\Gamma}(b,2|1,a)$.
The 4-15th terms in \eq{kernel} are of the form 
${\dot{\cal G}}^{(1)}{\ddGcal}_{\phantom{aa}}{\dot{\cal G}}^{(1)}$, 
and these are made of the form  
$(\theta_a+\theta_b)(1+\theta_a\theta_b)(\theta_a+\theta_b)$, 
whichever ${\cal G}^{(1)}$ we use. These terms are hence zero for 
${\hat\Gamma}_2^{(1)}(b,2|1,a)$. Since the remaining first  
three terms are composed of only ${\ddGcal}_{\phantom{aa}}$, 
the result is independent of the choice of ${\cal G}^{(1)}$
\beq
{\hat\Gamma}_2^{(1)}(b,2|1,a)
=(D-1)\eps_1\cdot\eps_2\int d\tau_1d\tau_2 
e^{k_1\cdot k_2 G_B^{(1)}(1,2)} (G_F^{12})^2{\cal N}_1^B\Bigr|_
{a\ra1,b\ra2}.       \label{nonzero2point}
\eeq

Next, let us show that 
\beq
           {\hat\Gamma}_2^{(1)}(2,1)=0.
\eeq
In the case of $G^{(1)}$, taking the limit $1\ra2$ in \eq{kernel}, 
it is easy to see this equality because of the properties 
\beqa
& \lim_{\tau_1\ra\tau_2}\theta_1\theta_2
                 {\ddG}_{\phantom{aa}}(1,2)=0, \\
& \lim_{\tau_1\ra\tau_2}{\dot G}^{(1)}(1,2)=0,    \label{limg12}
\eeqa
which mean ${\ddG}_{\phantom{aa}}(j,i)$ and ${\dot G}^{(1)}(j,i)$ 
can be dropped in the pinch situation $j\ra i$. In the case of 
${\tilde G}^{(1)}$, the first five terms in \eq{kernel} vanish 
from the same reason, because ${\ddGcal}_{\phantom{aa}}$ is 
independent of the choice of ${\cal G}^{(1)}$. 
The 6,7,8,9 and 10th terms cancel the 11,15,13,12 and 14th terms 
respectively. Therefore all 15 terms vanish, and the proof ends.

\section{Three-point function}
\setcounter{equation}{0}
\indent 

{}~For 3-point functions, we have to start with the following 8-fold 
correlator 
\beqa
{\hat\Gamma}_3^{(1)}&=&\int d\tau_a d\tau_b d\tau_1 d\tau_2 d\tau_3
    d\theta_a d\theta_b d\theta_1 d\theta_2 d\theta_3{\cal N}_1 \nn\\
& &<{\cal D}X_a^\sigma {\cal D}X_b^\delta
    {\cal D}X_1^\mu {\cal D}X_2^\nu {\cal D}X_3^\rho 
e^{ik_1\cdot X_1} e^{ik_2\cdot X_2}e^{ik_3\cdot X_3}>_{(1)}
g_{\sigma\delta}\eps^1_\mu \eps^2_\nu \eps^3_\rho.
\eeqa
Let us consider the pinch situation $2\ra1$. In this section, we 
only discuss the $G^{(1)}$ case which makes equations simpler owing 
to \eq{limg12}. Although the Wick contraction creates 68 terms, 
these are reduced to 30 terms by taking the pinch limit $2\ra1$. 
Further, using the interchange symmetry of $a\leftrightarrow b$, 
the number of terms becomes 15. Among them, the following 6 terms 
turn out to be independent
\beqa
I_1&=& {\ddG}_{ab}{\ddG}_{13}{\dot G}^{(1)}_{13},
\qquad 
I_2= {\ddG}_{a1}{\ddG}_{b3}{\dot G}^{(1)}_{13},\\
I_3&=& {\ddG}_{a1}{\ddG}_{13}{\dot G}^{(1)}_{b1},
\qquad
I_4= {\ddG}_{a1}{\ddG}_{13}{\dot G}^{(1)}_{b3},\\
I_5&=& {\ddG}_{13}{\dot G}^{(1)}_{a1}{\dot G}^{(1)}_{b1}
       {\dot G}^{(1)}_{13},
\qquad
I_6= {\ddG}_{a1}{\dot G}^{(1)}_{b1}{\dot G}^{(1)}_{13}
       {\dot G}^{(1)}_{31}.
\eeqa
After all, using $k_3=-(k_1+k_2)$, the pinched function 
${\hat\Gamma}_3^{(1)}(2,1)$ is expressed as
\beqa
{\hat\Gamma}^{(1)}_3(2,1) &=&2i \int d\tau_a d\tau_b d\tau_1 d\tau_2 
d\tau_3 d\theta_a d\theta_b d\theta_1 d\theta_2 d\theta_3 \nn\\
& & {\cal N}_1 \theta_2\theta_1\delta(\tau_2-\tau_1) 
\exp[\sum_{i<j}^3k_i\cdot k_j G^{(1)}(i,j)]   \nn\\
& &\times  [ ({D\over2}I_1 -I_2+I_3-I_4-k_1\cdot k_2 I_5)
(\eps_1\cdot\eps_3 \eps_2\cdot k_3 
                 - \eps_2\cdot\eps_3 \eps_1\cdot k_3)  \nn \\
& &\hskip15pt+ I_6\eps_1\cdot k_3(\eps_2\cdot k_1 \eps_3\cdot k_2
                         +\eps_2\cdot k_2 \eps_3\cdot k_1)   
\label{threepointI} \\
& &\hskip15pt- I_6\eps_2\cdot k_3(\eps_1\cdot k_1 \eps_3\cdot k_2
                         +\eps_1\cdot k_2 \eps_3\cdot k_1) ]. \nn
\eeqa
Let us consider the integrations on $\tau_2$ and all $\theta_i$. 
To this end, pick up only $\theta_a\theta_b\theta_3$ terms from the 
integrand ${\cal N}_1I_i\exp[\sum_{i<j} k_i\cdot k_j G^{(1)}(i,j)]$, 
and note 
\beq
G^{(1)}(1,3)=G^{(1)}_B(1,3)+{1\over2} g_{ab}(1,3)
(\theta_a\theta_3 G_F^{a3}-\theta_b\theta_3 G_F^{b3})
+{\cal O}(\theta_1,\theta_2),
\eeq
where
\beq
g_{ab}(i,j)=
{G_B^{ia}-G_B^{ib}-G_B^{aj}+G_B^{bj}\over{\bar T}+G_B^{ab}}.
\eeq 
This, with $\theta_a$ and $\theta_b$ terms of $I_i$, actually 
contributes to $\theta_a\theta_b\theta_3$ terms which are 
proportional to $k_3^2$. Also, a $\theta_a\theta_b$ term from 
${\cal N}_1$ contributes with $\theta_3$ terms of $I_i$. 
Then integrating $\theta_a$, $\theta_b$ and $\theta_3$,
\beq
J_i=\int d\theta_3 d\theta_b d\theta_a 
        I_i\exp[-k_3^2 G_B^{(1)}(1,3)], 
\eeq
we obtain the following values of $J_i$, 
\beqa
J_1 &=& (G_F^{13})^2({\ddot G}_B^{ab}+{1\over2}
{({\dot G}_B^{ab})^2-(G_F^{ab})^2 \over {\bar T}+ G_B^{ab}})
-{D\over2}{(G_F^{ab})^2(G_F^{13})^2\over{\bar T}+G_B^{ab}} \nn\\
& &+{1\over4}k_3^2 g_{ab}^2(1,3)(G_F^{1b}G_F^{a3}-G_F^{1a}G_F^{b3})
G_F^{ab}G_F^{13},
\eeqa
\beqa
J_2+J_4-J_3 &=& {1\over2}(G_F^{a1})^2{\ddot G}_B^{b3}g_{ab}(1,3) 
       +{1\over2}(G_F^{a1})^2{ ({\dot G}_B^{a3}-{\dot G}_B^{b3})
       ({\dot G}_B^{b1}-{\dot G}_B^{b3})\over {\bar T}+G_B^{ab}}\nn\\
&+ &{1\over2}k_3^2G_F^{a3}G_F^{a1}G_F^{13}({\dot G}_B^{b1}
       -{\dot G}_B^{b3})g_{ab}(1,3)
+{1\over4}k_3^2 g^2_{ab}(1,3)(G_F^{ab}G_F^{b3}-
{\dot G}_B^{ab}G_F^{a3})G_F^{a1}G_F^{13} \nn \\
&+ &{1\over4}k_3^2g_{ab}^2(1,3) (G_F^{1b}G_F^{a3}-
G_F^{1a}G_F^{b3})G_F^{a1}G_F^{b3}-D{G_F^{a1}G_F^{13}G_F^{3b}G_F^{ba}
\over{\bar T}+G_B^{ab}},
\eeqa
\beqa
J_5 &=& -(G_F^{13})^2 ( {\dot G}_B^{a1}-{1\over2}{\dot G}_B^{ab}
         { G_B^{ab}+G_B^{a1}-G_B^{b1}\over {\bar T}+G_B^{ab}} ) 
( {\dot G}_B^{b1}-{1\over2}{\dot G}_B^{ab}
{ G_B^{ab}+G_B^{b1}-G_B^{a1}\over {\bar T}+G_B^{ab}} )  \nn \\
& &-{1\over4}(G_F^{13}G_F^{ab})^2
{ (G_B^{ab}+G_B^{a1}-G_B^{b1})(G_B^{ab}+G_B^{b1}-G_B^{a1})
\over ({\bar T}+G_B^{ab})^2 }, 
\eeqa
\beqa
J_6&=& {1\over4}G_F^{a1}G_F^{ba}{G_B^{ab}-G_B^{a1}+G_B^{b1}\over
{\bar T}+G_B^{ab}}g_{ab}(1,3)
\left( G_F^{1b}\{ -{\dot G}_B^{31}+{1\over2}(
{\dot G}_B^{3a}-{\dot G}_B^{3b})g_{ab}(3,1)\} 
+G_F^{13}G_F^{3b}\right)                      \nn\\
& &-{1\over2}G_F^{a1}(-{\dot G}_B^{b1}+{1\over2}{\dot G}_B^{ba}
{G_B^{ab}-G_B^{a1}+G_B^{b1}\over{\bar T}+G_B^{ab}} )g_{ab}(1,3)\nn\\
& &\hskip80pt\times
\left(G_F^{13}G_F^{3a} + G_F^{1a}\{-{\dot G}_B^{31}+
{1\over2}({\dot G}_B^{3a}-{\dot G}^{3b})g_{ab}(3,1)\}\right),
\eeqa
which should be finally integrated in the expression
\beqa
{\hat\Gamma}^{(1)}_3(2,1) &=& -2i\int d\tau_a d\tau_b d\tau_1 d\tau_3
\exp[ k_3\cdot(k_1+k_2)G_B^{(1)}(\tau_1,\tau_3)]{\cal N}_1^B \nn\\ 
& &\times [\, ({D\over2}J_1 -J_2+J_3-J_4-k_1\cdot k_2 J_5)
(\eps_1\cdot\eps_3 \eps_2\cdot k_3 
                      - \eps_2\cdot\eps_3 \eps_1\cdot k_3)\nn \\
& &\hskip20pt+ J_6 \eps_1\cdot k_3(\eps_2\cdot k_1 \eps_3\cdot k_2
                         +\eps_2\cdot k_2 \eps_3\cdot k_1) \\
& &\hskip20pt- J_6\eps_2\cdot k_3(\eps_1\cdot k_1 \eps_3\cdot k_2
                         +\eps_1\cdot k_2 \eps_3\cdot k_1) \,]. \nn
\eeqa

Similarly, the following double-pinch function can be calculated 
\beq
{\hat\Gamma}^{(1)}_3(b,2|1,a)\sim{\hat\Gamma}^{(1)}_3
\theta_b\theta_2\theta_1\theta_a
\delta(\tau_b-\tau_2)\delta(\tau_1-\tau_a).
\eeq
The pinch $\delta$-functions $a\ra1$ and $b\ra2$ remove 34 terms 
(among the 68 terms). 
In addition, the following formula
\beq
{\cal D}_i G^{(1)}(i,j) =0 
+{\cal O}(\theta_1,\theta_2,\theta_a,\theta_b)
\qquad\mbox{$i,j \not=3$}
\eeq
removes further 24 terms. Gathering remaining 10 terms, we obtain
\beqa
{\hat\Gamma}^{(1)}_3(b,2|1,a) &=& -i\int d\tau_1 d\tau_2 d\tau_3
{\cal N}_1^B\exp[\sum_{i<j}k_j\cdot k_i G^{(1)}_B(\tau_j,\tau_i)]
\Bigr|_{a\ra1,b\ra2} \nn\\
& &\times [\,(D-1)\eps_1\cdot\eps_2\eps_3\cdot k_1(G_F^{12})^2
\{\, -{\dot G}_B^{31}+{1\over2}({\dot G}_B^{31}-{\dot G}_B^{32})
g_{ab}(3,1)\,\}  \nn\\
& &\hskip10pt + (D-1)\eps_1\cdot\eps_2\eps_3\cdot k_2(G_F^{12})^2
\{\, -{\dot G}_B^{32}+{1\over2}({\dot G}_B^{31}-{\dot G}_B^{32})
g_{ab}(3,2)\,\}  \nn \\ 
& &\hskip10pt + (D-2)G_F^{12}G_F^{23}G_F^{31}(\eps_1\cdot\eps_3 
\eps_2\cdot k_3 - \eps_2\cdot\eps_3 \eps_1\cdot k_3)\,]. 
\eeqa

\section{Pinched $N$-point functions}
\setcounter{equation}{0}
\indent 

In previous sections, we have calculated one-, two- and three-point 
functions directly, {\it i.e.} according to the 
definition \eq{pinchdef}. This kind of calculations becomes more 
complicated as increasing the number of external legs. However, 
there must be a simple method to evaluate two-loop $N$-point 
functions, if we start from known one-loop $(N+2)$-point functions. 
In this section, let us consider this possibility for a while.

The un-pinched two-loop function eq.\eq{twoloop} can be written 
in a similar form as the one-loop formula \eq{oneloopgamman} 
\beq
{\hat\Gamma}_N^{(1)}=\oint[dX]e^{-S_1} V_bV_a\prod_{n=1}^N V_n
\Bigr|_{k_a=k_b=0, \eps_a^\mu\eps_b^\nu=g^{\mu\nu}},
\label{newform}
\eeq
where
\beq
S_1 = -{(X_a-X_b)^2\over4{\bar T}}
       -{1\over4}\int_0^T d\tau d\theta X{\cal D}^3X.
\eeq
This suggests that we have only to (i) evaluate one-loop 
$(N+2)$-point function $<V_bV_a\prod V_n>$, (ii) then substitute 
one-loop quantities by two-loop ones 
\beq
{\cal N}\quad\ra\quad{\cal N}_1, \qquad 
G_B(\tau_1,\tau_2)\quad\ra\quad G_B^{(1)}(\tau_1,\tau_2),
\label{reduction1} 
\eeq
with
\beq
\eps_a^\mu \eps_b^\nu\quad\ra\quad g^{\mu\nu},\qquad
k_a,k_b\quad\ra\quad 0,  \label{reduction2}
\eeq
where $a$ and $b$ are supposed to be the leg labels which will be 
joined by a propagator insertion to make another loop. The rules 
\eq{reduction2} coincide with those shown in \cite{sqed}. However, 
note that $G_B^{(1)}$ should be the combination of $G_B$ defined 
not in \eq{Gofss} but in \eq{Gfromstring}, in accordance with 
the fact $G_B(\tau_i,\tau_i)=0$. 

We can easily check the validity of the above replacements up to 
between one-loop pinched 4-point functions and two-loop pinched 
2-point functions. The first simple example is between the one-loop 
${\hat\Gamma}_3(2,1)$ and two-loop ${\hat\Gamma}^{(1)}_1(1,a)$ 
functions. Choosing the joining label set $(a,b)$ to be $(3,1)$ in 
eq.\eq{3pointpinch2} and setting $k_3=k_1=0$, 
$\eps_3^\mu\eps_1^\nu=g^{\mu\nu}$, we see that 
\beq
        {\hat\Gamma}_3(2,1)\quad\ra\quad 0,
\eeq
which coincides with eq.\eq{star} {\it i.e.} 
${\hat\Gamma}_1^{(1)}(1,a)$. Similarly, if we choose $(a,b)=(1,2)$ 
in eq.\eq{4pointpinch21} and $(a,b)=(3,4)$ in eq.\eq{4pointpinch21}, 
we verify the following map relations respectively 
\beqa
{\hat\Gamma}_4(2,1)\quad\ra\quad 0 &=& {\hat\Gamma}_2^{(1)}(b,a),\\
{\hat\Gamma}_4(2,1)\quad\ra\quad 0 &=& {\hat\Gamma}_2^{(1)}(2,1).
\eeqa
For a non-zero example, put $(a,b)=(4,1)$ in eq.\eq{4pointpinch4321}. 
Then we recover eq.\eq{nonzero2point} from the one-loop 4-point 
function 
\beqa
{\hat\Gamma}_4(4,3|2,1)\quad&\ra&\quad 
(D-1)\eps_3\cdot\eps_2\int d\tau_3 d\tau_2 {\cal N}_1^B 
(G_F^{32})^2 \exp[k_3\cdot k_2 G_B^{(1)}(\tau_3,\tau_2)],\nn\\
&=& {\hat\Gamma}_2^{(1)}(b,3|2,a),
\eeqa
where $\tau_a$ and $\tau_b$ in ${\cal N}_1^B$ must be replaced by 
$\tau_2$ and $\tau_3$.  

It is nontrivial whether \eq{reduction1} is further valid for the 
two-loop pinched 3-point functions obtained in previous section, 
since several contributions from $G^{(1)}$ and ${\cal N}_1$ exist 
as remarked there --- although the un-pinched function \eq{newform} 
clearly support the replacement \eq{reduction1}. 
For the $(m+1)$-loop cases of Schmidt-Schubert's type \cite{phi3}, 
\cite{qed}, we can easily generalize these replacements, 
using $m$ copies of \eq{reduction2} and introducing 
${\cal N}_m$ and $G_B^{(m)}$ in \eq{reduction1} 
(${\cal N}_m$ and $G_B^{(m)}$ are given in eq.(28) of \cite{phi3}, 
however note that the $G_B^{(m)}$ here should be the one which 
satisfies $G_B^{(m)}(x,x)=0$).

\section{Gluon loop}
\setcounter{equation}{0}
\indent 

Our pinch technique can be easily applied to gluon loop cases as 
well, once the un-pinched part of $N$-point function 
${\hat\Gamma}^{(1)}_N$ for gluon loop is written in a form of 
super-worldline correlator. Let us recall the difference between 
spinor and gluon loops at the one-loop level. In the gluon loop 
case, we have to change \cite{str}
\beq
S_0 \quad\ra\quad {\tilde S}_0={1\over4}{\dot x}^2 
                + \psi_+{\dot \psi}_-  +  C(\psi_+\psi_- -1),
\eeq
\beq
\oint[dX] \quad\ra\quad \lim_{C\ra\infty} 
\sum_{p=0}^1{1\over2}(-)^{p+1}\oint_{(p)}[dX],
\eeq
where $p=0$ ($p=1$) denotes (anti-)periodic worldline fermions, 
which are defined by $\psi=\psi_+ + \psi_-$ and satisfy
$\{\psi_+^\mu,\psi_-^\nu\}=g^{\mu\nu}$, $\{\psi_\pm, \psi_\pm\}=0$. 
The above action inevitably changes the fermionic worldline Green 
function $G_F$ into 
\beqa
G_F^{(1)}(\tau_1,\tau_2) &=& 2{\rm sign}(\tau_1-\tau_2)
            e^{-CT/2}{\rm cosh}({CT\over2}-C|\tau_1-\tau_2|), \\       
G_F^{(0)}(\tau_1,\tau_2) &=& 2{\rm sign}(\tau_1-\tau_2)
            e^{-CT/2}{\rm sinh}({CT\over2}-C|\tau_1-\tau_2|) .
\eeqa

Taking account of these modifications, `gluon' two-loop functions 
may be written as 
\beq
{\hat\Gamma}_N^{(1)}=\int d\tau_a d\tau_b d\theta_a d\theta_b 
\lim_{C\ra\infty}\sum_{p=0}^1{1\over2}(-)^{p+1} 
{\cal N}_1^{(p)}<{\cal D}X_b\cdot{\cal D}X_a\prod_{n=1}^N
V_n>^p_{(1)},    \label{gluontwoloop}
\eeq
where
\beq 
{\cal N}_1^{(p)}=\oint_{(p)}[dX]e^{-{\tilde S}_0}
\exp[-{(X_a-X_b)^2\over{\bar T}}] =    
Z_p (4\pi T)^{-D\over2}(1+{1\over{\bar T}}G(a,b))^{-D\over2},
\eeq
\beq
Z_p =e^{CT}(1 + (-)^{p+1} e^{-CT})^4, 
\eeq
and 
\beq
<X_1X_2>_{(1)}^p = {1\over {\cal N}_1^{(p)} } \oint_{(p)}[dX]
e^{-{\tilde S}_0}\exp[-{(X_a-X_b)^2\over{\bar T}}]X_1X_2 .
\eeq    
Note that $G(i,j)$ is now modified 
\beq
G(i,j) = G_B(\tau_i,\tau_j) 
        + 2\theta_1\theta_2 G_F^{(p)}(\tau_1,\tau_2). 
\eeq
Similarly as the spinor loop case, 
pinched functions of \eq{gluontwoloop} can be obtained through 
inserting pinch $\delta$-functions 
$\theta_j\theta_i\delta(\tau_j-\tau_i)$ etc. 
Now, the simplest example is the 
two-point function ${\hat\Gamma}_2^{(1)}(b,2|1,a)$
\beqa
{\hat\Gamma}_2^{(1)}(b,2|1,a)
&=&(D-1)\eps_1\cdot\eps_2\int d\tau_1d\tau_2
e^{k_1\cdot k_2 G_B^{(1)}(1,2)}
\lim_{C\ra\infty}\sum_{p=0}^1{1\over2}(-)^{p+1}{\cal N}_1^p 
[2G_F^{(p)}(\tau_1,\tau_2)]^2\nn\\
&=& 8 (D-1)\eps_1\cdot\eps_2\int d\tau_1d\tau_2
e^{k_1\cdot k_2 G_B^{(1)}(1,2)}
(4\pi T)^{-D\over2}(1+{1\over{\bar T}} G_B^{12})^{-D\over2}.
\label{gluetwo}
\eeqa
Other pinched functions can be obtained from the pinched functions 
${\hat\Gamma}_N^{(1)}$ of spinor loop through the replacement rule 
which was developed in one-loop studies \cite{str}. Namely, replace 
the {\it single} $G_F$ chain appeared in the ${\hat\Gamma}_N^{(1)}$ 
of {\it spinor loop}
\beq
[\prod_{k=1}^d G_F^{i_{k+1},i_k}] = \left\{ \begin{array}{ll}
   -2^d & \mbox{if $\tau_{max}=
           \tau_{i_d}>\tau_{i_{d-1}}>\cdots>\tau_{i_2}>\tau_{i_1}
                              =\tau_{min}$} \\
   -(-2)^d & \mbox{if $\tau_{max}=
           \tau_{i_d}>\tau_{i_1}>\tau_{i_2}>\cdots>\tau_{i_{d-1}}
                                 =\tau_{min}$} \\
   -8 & \mbox{if $d=2$} \\
   0 & \mbox{otherwise.} \\
\end{array} \right.
\eeq
Of course, this rule recovers the second line of eq.\eq{gluetwo}. 
As seen from the above argument, the gluon loop case is almost same 
as the spinor loop case. An important point is that every pinched 
function can be easily evaluated by inserting a possible number of 
the $\delta$-function $\theta_j\theta_i\delta(\tau_j-\tau_i)$ 
as long as the (un-pinched) $N$-point function is expressed in a
form of super-worldline correlator. This enables us to calculate 
any pinch between external and internal gluon lines.

\section{Conclusion}
\setcounter{equation}{0}
\indent 

We have, using the super worldline formalism, reformulated 
Strassler's method to evaluate pinched $N$-point functions associated 
to the quadratic terms of field strength in non-abelian gauge theory. 
It is much convenient that the pinched gluon vertex operator has been 
written in the elegant superfield expression which is given by the 
product of two gluon vertex operators where the pinch 
$\delta$-function is inserted. Owing to this result, we have only to 
insert the pinch $\delta$-function 
$\theta_j\theta_i\delta(\tau_j-\tau_i)$ into the un-pinched $N$-point 
function ${\hat\Gamma}_N$, in order to obtain the pinched function 
${\hat\Gamma}_N(j,i)$. After that, one can put an appropriate color 
factor. This method makes calculation simpler and more transparent 
than the original pinch method of Strassler. Furthermore, 
our method can be applied straightforwardly to two-loop diagrams 
(although the present cases are nothing but the QED corrections), 
and we have derived various formulae on the two-loop pinched $N$-point 
functions. In particular, the pinch between internal and external 
gluon has become calculable in terms of the $\delta$-function 
insertion. 

Of course, there exist several unsolved problems as well. First of 
all, the direct proof of coincidence of our two-loop $N$-point 
functions with Feynman rule results is difficult and non-trivial 
at the present expressions, since our expressions of $N$-point 
functions are extremely different from standard ones. For example, 
it is hard to guess the origin of two-loop Green function $G_B^{(1)}$ 
in the corresponding Feynman rule calculation. A reasonable 
comparison could be done after all (or some of) integrations w.r.t. 
$\tau$ variables, like done in \cite{qed} --- where they performed 
every $\tau$ integration using the Fock-Schwinger gauge and 
extracted a relevant divergent constant to the QED 
$\beta$-function. This kind of analysis would seem to be the 
only one that we can do at the present level of our techniques. 

Second one is the following. We have considered the equivalence 
between two different Green functions in the situations of pinched 
one- and two-point functions, whereas it is still unclear in the 
un-pinched function cases. In these pinched situations, all the 
$\tau$-integrations were easy because of the $\delta$-function 
insertions, which reduce the number of non-trivial integrations 
on $\tau$ variables. The equivalence in un-pinched functions 
would become clear after performing all of integrations on $\tau_a$, 
$\tau_b$ and $\tau_n$, although we did not analyze the cases. 
Otherwise, there should be found a vanishing integral formula to 
compensate the difference of Green functions.

Thirdly, we have not considered any insertions of vertex operators 
into the inserted `gluon' propagator line. This kind of vertex operator 
insertions occurs also in more general situation like in 
Schmidt-Schubert type multi-loop diagrams \cite{phi3},\cite{qed}. 
In scalar $\phi^3$-theory, worldline Green functions for such 
vertex insertions are already found in \cite{HTS},\cite{RS}. 
As easily expected from the scalar theory analysis, these multi-loop 
cases have further three types of worldline Green functions, and the 
$N$-point function exponent of these types should be given by the 
sum of these possible vertex insertions. Our pinch formalism 
presented here could be basically applied to these complicated 
multi-loop diagrams. However, having various types of worldline 
Green functions, we do not know which one should be used in Wick 
contractions when evaluating non-exponent parts 
$K_{\mu\nu\dots\rho}$. Without solving this problem, we will 
not be able to go ahead for further complicated multi-loop 
diagram analyses. 

Finally, another consistency check should be pursued from the 
string theoretical approach. String theories offer us a reasonable 
input to particle theory through the infinite limit of 
string tension. Since a particle diagram corresponds to a corner of 
moduli space, we can start from a universal expression of 
world-sheet Green function \cite{Mart}. This might solve one of the 
above problems. Also, it is interesting to note that the pinch 
prescription presented here is very similar to the one 
developed in superstring theory \cite{AT}.  

Although we have not yet arrived at any familiar result obtained by 
the standard (Feynman rule) calculation, we believe that our pinch 
formulation and formulae will be useful for further understanding 
and development of Bern-Kosower-like rules for multi-loop 
scattering amplitudes in Yang-Mills theory.   

\vspace{0.3cm}
\noindent
{\bf Acknowledgments}\par
The author would like to thank M.G. Schmidt for useful 
suggestion and comments.

\vspace{0.3cm}
\noindent
{\bf Note added:} In this paper, we concentrated on the photon 
propagator insertion. After submitting this work, an exact path 
integral expression of gluon propagator has been proposed in 
\cite{gl}. Together with that, one can perform some 
calculations of pinch contribution from the internal gluon line. 

%
\end{document}